\documentclass[twocolumn,groupedaddress,superscriptaddress,nofootinbib]{revtex4}
\usepackage{graphicx,amsmath,amssymb,amsbsy,txfonts, float}
\usepackage{subfigure,hyperref,bbm,times}
\usepackage[T1]{fontenc}
\usepackage{braket}
\usepackage{epsfig}
\usepackage{color}
\usepackage{graphicx}
\usepackage{dcolumn}
\usepackage{bm}

\providecommand{\openone}{\leavevmode\hbox{\small1\kern-3.8pt\normalsize1}}

\usepackage{soul}

\hypersetup{
   colorlinks=true,
   linkcolor=blue,
}
\graphicspath{{figure/}}
\begin{document}

\title{Robust entanglement preparation against noise by controlling spatial indistinguishability}

\author{Farzam Nosrati}
\affiliation{Dipartimento di Ingegneria, Universit\`{a} di Palermo, Viale delle Scienze, Edificio 9, 90128 Palermo, Italy}
\affiliation{INRS-EMT, 1650 Boulevard Lionel-Boulet, Varennes, Qu\'{e}bec J3X 1S2, Canada}

\author{Alessia Castellini}
\affiliation{Dipartimento di Fisica e Chimica - Emilio Segr\`e, Universit\`a di Palermo, via Archirafi 36, 90123 Palermo, Italy}

\author{Giuseppe Compagno}
\affiliation{Dipartimento di Fisica e Chimica - Emilio Segr\`e, Universit\`a di Palermo, via Archirafi 36, 90123 Palermo, Italy}

\author{Rosario Lo Franco}
\email{rosario.lofranco@unipa.it}
\affiliation{Dipartimento di Fisica e Chimica - Emilio Segr\`e, Universit\`a di Palermo, via Archirafi 36, 90123 Palermo, Italy}
\affiliation{Dipartimento di Ingegneria, Universit\`{a} di Palermo, Viale delle Scienze, Edificio 6, 90128 Palermo, Italy}

\begin{abstract}
Initialization of composite quantum systems into highly entangled states is usually a must to enable their use for quantum technologies. However, unavoidable noise in the preparation stage makes the system state mixed, hindering this goal. Here we address this problem in the context of identical particle systems within the operational framework of spatially localized operations and classical communication (sLOCC). We define the entanglement of formation for an arbitrary state of two identical qubits. We then introduce an entropic measure of spatial indistinguishability as an information resource. Thanks to these tools we find that spatial indistinguishability, even partial, can be a property shielding nonlocal entanglement from preparation noise, independently of the exact shape of spatial wave functions. These results prove quantum indistinguishability is an inherent control for noise-free entanglement generation.
\end{abstract}

\maketitle

\section{Introduction}

The discovery and utilization of purely quantum resources is an ongoing issue for basic research in quantum mechanics and quantum information processing \cite{trabesinger2017quantum,ladd2010quantum}. 
Processes of quantum metrology \cite{giovannetti2006quantum}, quantum key distribution \cite{ekert1991quantum}, teleportation \cite{PhysRevLett.70.1895} or quantum sensing \cite{quantumsensingRMP} essentially rely on the entanglement feature \cite{audretsch,horo2009}. Unfortunately, entanglement is fragile due to the inevitable interaction between system and surrounding environment already in the initial stage of pure state preparation, making the state mixed \cite{chuang2010,aolitareview}. As a result, protecting entanglement from unavoidable noises remains a main objective for quantum-enhanced technology \cite{lofrancoreview}. 

Many-body quantum networks usually employ identical quantum subsystems (e.g., qubits) as building blocks \cite{bloch2008many,anderlini2007controlled,wang2016experimental,
cronin2009ad,crespi2015particle,PhysRevLett.116.116801,martinisfermions,RevModPhys.90.035006}. Characterizing peculiar features linked to particle indistinguishability in composite systems assumes importance from both the fundamental and technological points of view. 
Discriminating between indistinguishable and distinguishable particles has always been a big challenge for which different theoretical \cite{PhysRevLett.113.020502,Aolita2015,Bentivegna_2016,Dittel_2017} and experimental \cite{Spagnolo2014,bentivegna,Crespi2016,PhysRevX.9.011013,Giordani2018} techniques have been suggested. 
Recently, particle identity and statistics have been shown to be a resource \cite{PhysRevLett.118.153602, PhysRevLett.118.153603,PhysRevLett.88.187903,PhysRevA.68.052309,benatti2014NJP,PhysRevLett.120.240403} and experiments which witness its presence have been performed \cite{PhysRevLett.122.063602}.  
One aspect that remains unexplored is how the continuous control of the spatial configurations of one-particle wave functions, ruling the degree of indistinguishability of the particles, influences noisy entangled state preparation. 
Moreover, a measure of the degree of indistinguishability lacks. 

Pursuing this study requires an entanglement quantifier for an arbitrary (mixed) state of the system with tunable spatial indistinguishability. 
It is desirable that this quantifier is defined within a suitable operational framework reproducible in the laboratory. The natural approach to this aim is the recent experimentally-friendly framework based on spatially localized operations and classical communication (sLOCC), which encompasses entanglement under generic spatial overlap configurations \cite{PhysRevLett.120.240403}. This approach has been shown to also enable remote entanglement \cite{castelliniPRA19,PhysRevA.96.022319} and quantum coherence \cite{castellini2019indistinguishability}.

Here we adopt the sLOCC framework to unveil further fundamental traits of composite quantum systems. We first define the entanglement of formation for an arbitrary state of two indistinguishable qubits (bosons or fermions). We then introduce the degree of indistinguishability as an entropic measure of information, tunable by the shapes of spatial wave functions. We finally apply these tools to a situation of experimental interest, that is noisy entangled state preparation. We find spatial indistinguishability can act as a tailored property protecting entanglement generation against noise.

\section{Results}

\textbf{sLOCC-based entanglement of formation of an arbitrary state of two identical qubits.} We first focus on the quantification of entanglement for an arbitrary state (pure or mixed) of identical particles.  For identical particles we in general mean identical constituents of a composite system.
In quantum mechanics identical particles are not individually addressable, as are instead non-identical (distinguishable) particles, so that specific approaches are needed to treat their collective properties \cite{tichy2011essential,bose2013,balachandranPRL,plenio2014PRL,benattiOSID2017,
LoFranco2016,compagnoRSA,duzzioniPRA}.
Our goal is accomplished by straightforwardly redefining the entanglement of formation known for distinguishable particles \cite{PhysRevLett.80.2245} to the case of indistinguishable particles, thanks to the sLOCC framework \cite{PhysRevLett.120.240403}.

The separability criterion in the standard theory of entanglement for distinguishable particles \cite{horo2009,PhysRevLett.80.2245} maintains its validity also for a state of indistinguishable particles once it has been projected by sLOCC onto a subspace of two separated locations $\mathcal{L}$ and $\mathcal{R}$. In fact, after the measurement, the particles are individually addressable into these regions and the criteria known for distinguishable particles can be adopted \cite{PhysRevLett.120.240403,castellini2019indistinguishability}.

Consider two identical qubits, with spatial wave functions $\psi_1$ and $\psi_2$, for which one desires to characterize the entanglement between the pseudospins between the separated operational regions. States of the system can be expressed by the elementary-state basis 
$\{\ket{\psi_1\sigma_1,\psi_2\sigma_2},\ \sigma_1,\sigma_2=\uparrow, \downarrow\}$, expressed in the no-label particle-based approach \cite{LoFranco2016,compagnoRSA} where fermions and bosons are treated on the same footing. The density matrix of an arbitrary state of the two identical qubits can be written as
\begin{equation}\label{Eq:6}
\rho=\sum_{\sigma_1,\sigma_2,\sigma'_1,\sigma'_2=\downarrow,\uparrow}p_{\sigma_1\sigma_2}^{\sigma'_1,\sigma'_2}\ket{\psi_1 \sigma_1,\psi_2 \sigma_2}\bra{\psi_1 \sigma'_1,\psi_2 \sigma'_2}/\mathcal{N},
\end{equation}
where $\mathcal{N}$ is a normalization constant.
Projecting $\rho$ onto the (operational) subspace spanned by the computational basis $\mathcal{B}_\mathrm{LR}=\{\ket{\mathrm{L}\uparrow,\mathrm{R}\uparrow}, \ket{\mathrm{L}\uparrow,\mathrm{R}\downarrow}, \ket{\mathrm{L}\downarrow,\mathrm{R}\uparrow}, \ket{\mathrm{L}\downarrow,\mathrm{R}\downarrow}\}$ by the projector
\begin{equation}\label{PiLR}
\Pi_{\mathrm{LR}}^{(2)}=\sum_{\tau_1,\tau_2=\uparrow,\downarrow}\ket{\mathrm{L}\tau_1,\mathrm{R}\tau_2}\bra{\mathrm{L}\tau_1,\mathrm{R}\tau_2},
\end{equation}
one gets the distributed resource state
\begin{equation}\label{rhoLR}
\rho_\mathrm{LR}=\Pi_\mathrm{LR}^{(2)}\rho\Pi_\mathrm{LR}^{(2)}/\mathrm{Tr}(\Pi_\mathrm{LR}^{(2)}\rho), 
\end{equation}
with probability $P_\mathrm{LR}=\mathrm{Tr}(\Pi_\mathrm{LR}^{(2)}\rho)$. We call $P_\mathrm{LR}$ sLOCC probability since it is related to the post-selection procedure to find one particle in $\mathcal{L}$ and one particle in $\mathcal{R}$. The state $\rho_\mathrm{LR}$ is then exploitable for quantum information tasks by addressing the individual pseudospins in the separated regions $\mathcal{L}$ and $\mathcal{R}$, which represent the nodes of a quantum network. The state $\rho_\mathrm{LR}$ can be in fact remotely entangled in the pseudospins and constitute the distributed resource state. The trace operation is clearly performed in the LR-subspace (see appendices \ref{AppA} and \ref{AppC}).
The state $\rho_\mathrm{LR}$, containing one particle in $\mathcal{L}$ and one particle in $\mathcal{R}$, can be diagonalized as 
$\rho_\textrm{LR}=\sum_{i}p_{i}\ket{\Psi_i^{\mathrm{LR}}}\bra{\Psi_i^{\mathrm{LR}}}$,
where $p_{i}$ is the weight of each pure state 
$\ket{\Psi_i^{\mathrm{LR}}}$ which is in general non-separable. 
Entanglement of formation of $\rho_\textrm{LR}$ is as usual \cite{PhysRevA.54.3824} $E_f(\rho_\textrm{LR})=\min \sum_{i}p_{i} E(\ket{\Psi_i^{\mathrm{LR}}})$, where minimization occurs over all the decompositions of $\rho_\mathrm{LR}$ and $E(\Psi_i^{\mathrm{LR}})$ is the entanglement of the pure state $\ket{\Psi_i^{\mathrm{LR}}}$. 
We thus define the operational entanglement $E_\mathrm{LR}(\rho)$ of the original state $\rho$ obtained by sLOCC as the entanglement of formation of $\rho_\mathrm{LR}$, that is 
\begin{equation}
E_\mathrm{LR}(\rho):=E_f(\rho_\mathrm{LR}).
\end{equation}

We can conveniently quantify the entanglement of formation $E_f(\rho_\mathrm{LR})$ by the concurrence $C(\rho_\mathrm{LR})$, according to the well-known relation $E_f=h[(1+\sqrt{1-C^2})/2]$ \cite{PhysRevLett.78.5022,PhysRevLett.80.2245}, where $h(x)=-x\log_2x-(1-x)\log_2(1-x)$. The concurrence $C_\mathrm{LR}(\rho)$ in the sLOCC framework can be easily introduced by
\begin{equation}\label{CLR}
C_\mathrm{LR}(\rho):=C(\rho_\mathrm{LR})=\max \{0,\sqrt{\lambda_4}-\sqrt{\lambda_3}-\sqrt{\lambda_2}-\sqrt{\lambda_1}\},
\end{equation}
where $\lambda_i$ are the eigenvalues, in decreasing order, of the non-Hermitian matrix $R=\rho_\mathrm{LR}\Tilde{\rho}_\mathrm{LR}$, being $\Tilde{\rho}_\mathrm{LR}=\sigma^\mathrm{L}_{y}\otimes\sigma_{y}^\mathrm{R} 
\rho_\mathrm{LR}^{*}\sigma_{y}^\mathrm{L}\otimes\sigma_{y}^\mathrm{R}$
with localized Pauli matrices $\sigma_{y}^\mathrm{L}=\ket{\mathrm{L}}\bra{\mathrm{L}}\otimes\sigma_{y}$, $\sigma_{y}^\mathrm{R}=\ket{\mathrm{R}}\bra{\mathrm{R}}\otimes\sigma_{y}$. 
The entanglement quantifier of $\rho$ so obtained contains all the information about spatial indistinguishability and statistics (bosons or fermions) of the particles.

\textbf{sLOCC-based entropic measure of indistinguishability.} In quantum mechanics, identical particles can be given the property of indistinguishability associated to a specific set of quantum measurements, being different from identity that is an intrinsic property of the system. With respect to the set of measurements, it seems natural to define a continuous degree of indistinguishability, which quantifies how much the measurement process can distinguish the particles. In this section, we deal with this aspect within sLOCC. For simplicity, the treatment is first presented for a two-particle pure state and then generalized to $N$-particle pure states. It is worth to mention that the framework is universal and also valid for mixed states.

\begin{figure*}[!t]
\begin{center}
\includegraphics[scale=0.147]{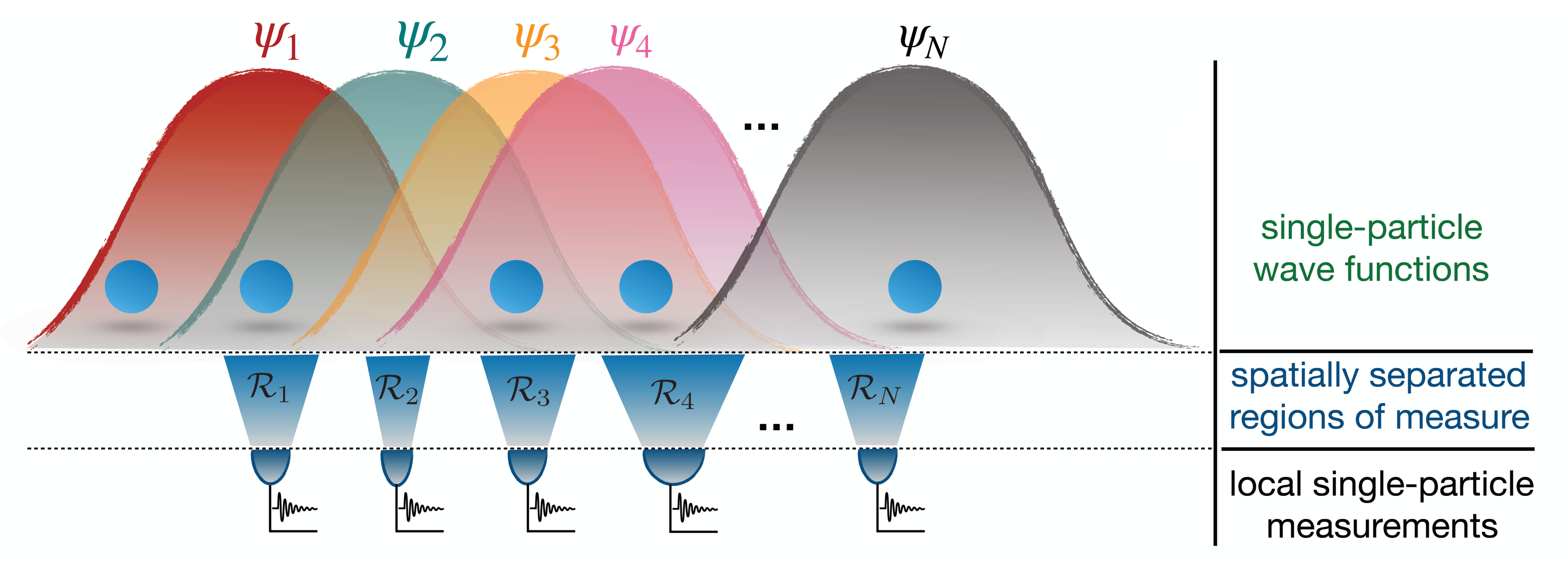}
\caption{\textbf{Projective measurements based on sLOCC.} Illustration of different single-particle spatial wave functions $\psi_i$ ($i=1,\ldots, N$) associated to $N$ identical particles in a generic spatial configuration. The amount of spatial indistinguishability of the particles can be defined by using spatially localized single-particle measurements in $N$ separated regions $\mathcal{R}_i$.}
\label{Figure1}
\end{center}
\end{figure*}

Let us consider an elementary pure state of two identical particles $\ket{\Psi^{(2)}}=\ket{\chi_1,\chi_2}$, where $\ket{\chi_i}$ is a generic one-particle state containing a set of commuting observables such as spatial wave function $\ket{\psi_i}$ and an internal degree of freedom $\ket{\sigma_i}$ (e.g., pseudospin with basis $\{\uparrow, \downarrow\}$). The $2$-particle state $\ket{\Psi^{(2)}}$ is thus  
\begin{gather}
\ket{\Psi^{(2)}}=\ket{\chi_1,\chi_2}=\ket{\psi_1\sigma_1,\psi_2\sigma_2}.
\end{gather}
In general, the degree of indistinguishability depends on both the quantum state and the measurement performed on the system. This means a given set of operations allows one to distinguish the particles while another set of operations does not. Let us narrow the analysis down to spatial indistinguishability within the sLOCC framework, linked to the incapability of distinguish which one of the two particles is found in each of the separated operational region. This framework thus leads to the concept of remote spatial indistinguishability of identical particles. The suitable class of measurements to this aim is represented by local counting of particles, leaving the pseudospins untouched. 
Inside this class, the joint projective measurement $\Pi_\mathrm{LR}^{(2)}$ defined in Eq. (\ref{PiLR}) represents the detection of one particle in $\mathcal{L}$ and of one particle in $\mathcal{R}$. We indicate with $P_{\mathrm{X\psi_i}}=|\langle \mathrm{X}|\psi_i\rangle|^2$ ($\mathrm{X}=\mathrm{L},\mathrm{R}$ and $i=1,2$) the probability of finding one particle in the region $\mathcal{X}$ ($\mathcal{X}=\mathcal{L},\mathcal{R}$) coming from $\ket{\psi_i}$. 
We then define the joint probabilities of the two possible events when one particle is detected in each region: (i) $\mathcal{P}_{12}=P_{\mathrm{L\psi_1}}P_{\mathrm{R}\psi_2}$ related to the event of finding a particle in $\mathcal{L}$ emerging from $\ket{\psi_1}$ and a particle in $\mathcal{R}$ emerging from $\ket{\psi_2}$, (ii) $\mathcal{P}_{21}=P_{\mathrm{L}\psi_2}P_{\mathrm{R}\psi_1}$ related to the vice versa. 
The amount of the no-which way information emerging from the outcomes of the joint sLOCC measurement $\Pi_\mathrm{LR}^{(2)}$ is a measure of the spatial indistinguishability of the particles in the state $\ket{\Psi^{(2)}}$.  
We thus use $\mathcal{Z}^{(2)}:=\mathrm{Tr}(\Pi_\mathrm{LR}^{(2)}\ket{\Psi^{(2)}}\bra{\Psi^{(2)}})=\mathcal{P}_{12}+\mathcal{P}_{21}$, that encloses the essence of this lack of information, to introduce the entropic measure of the degree of remote spatial indistinguishability
\begin{equation}
\mathcal{I_{\mathrm{LR}}}:=-\dfrac{\mathcal{P}_{12}}{\mathcal{Z}} \log_2 \dfrac{\mathcal{P}_{12}}{\mathcal{Z}}
-\dfrac{\mathcal{P}_{21}}{\mathcal{Z}}\log_2 \dfrac{\mathcal{P}_{21}}{\mathcal{Z}}.
\label{Indistinguishability}
\end{equation}
The entropic expression above naturally arises from the requirement of quantifying the \textit{no which-way information} associated to the uncertainty about the origin (spatial wave function) of the particle found in each of the operational regions. If particles do not spatially overlap in both remote regions, we have maximum information ($\mathcal{P}_{12}=1$, $\mathcal{P}_{21}=0$ or vice versa) and $\mathcal{I}_{\mathrm{LR}}=0$ (the particles can be distinguished by their spatial location). On the other hand, 
$\mathcal{I}_{\mathrm{LR}}=1$ when there is no information at all about each particle origin ($\mathcal{P}_{12}=\mathcal{P}_{21}$) and the particles are maximally overlapping in both regions. 
Notice that a given value of $\mathcal{I_{\mathrm{LR}}}$ corresponds to a class of different shapes of the single-particle spatial wave functions $\ket{\psi_i}$. 
Moreover, in an experiment which reconstructs the identical particle state by standard quantum tomography, the corresponding value of $\mathcal{I}_{\mathrm{LR}}$ can be indirectly obtained.

The above definition of the degree of spatial indistinguishability for two identical particles allows us to defining a more general degree of indistinguishability for $N$ identical particles. In general, $N$ different operational regions $\mathcal{R}_i$ ($i=1,\ldots,N$) are needed to quantify the indistinguishability of $N$ identical particles (see Fig.~\ref{Figure1}). 
Let us consider a $N$-identical particle elementary pure state $\ket{\Psi^{(N)}}=\ket{\chi_1,\chi_2,...,\chi_N}$, where $\ket{\chi_i}$ is the $i$-th single-particle state. Each $\ket{\chi_i}$ is characterized by the set of values $\chi_i=\chi_i^a, \chi_i^b,\ldots,\chi_i^{n}$ corresponding to a complete set of commuting observables $\hat{a}, \hat{b}, \ldots, \hat{n}$. For example, if $\hat{a}$ describes the spatial distribution of the single-particle states, $\chi_i^a$ is a spatial wavefunction $\psi_i$. 
To define a suitable class of measurements, we take the $N$-particle state
\begin{equation}\label{compactstate}
\ket{\alpha\beta}_N:=\ket{\alpha_1 \beta_1,\alpha_2\beta_2,\ldots,\alpha_N\beta_N},
\end{equation}
where the $i$-th single-particle state $\ket{\alpha_i\beta_i}$ is characterized by a subset $\hat{a},\hat{b},\ldots,\hat{j}$ of the $\hat{a},\hat{b},\ldots,\hat{n}$ commuting observables with eigenvalues $\alpha_i=\alpha_i^a,\alpha_i^b,\ldots,\alpha_i^j$, and by the remaining observables $\hat{k},\ldots,\hat{n}$ with eigenvalues $\beta_i=\beta_i^k,\ldots,\beta_i^n$. In the first member of Eq.~(\ref{compactstate}) we have set $\alpha:=\{\alpha_1,...,\alpha_N\}$ and $\beta:=\{\beta_1,...,\beta_N\}$. The $N$-particle projector on outcomes ($\alpha,\beta$) of the complete set of observables is $\Pi^{(N)}_{\alpha\beta}=\ket{\alpha\beta}_N\bra{\alpha\beta}$, while the projector on outcomes $\alpha$ of the partial set of observables is
\begin{equation}\label{projector}
\Pi^{(N)}_{\alpha}=\sum_{\beta}\Pi^{(N)}_{\alpha\beta}.
\end{equation}

Within the sLOCC framework, we can quantify to which extent particles in the state $\ket{\Psi^{(N)}}$ can be distinguished by knowing the results $\alpha$ of the local measurements described by $\Pi^{(N)}_{\alpha}$ of Eq. \eqref{projector}, considering that single-particle spatial wave functions $\{\psi_i\}$ can overlap (see Figure \ref{Figure1}). 
The (sLOCC) measurements have to satisfy the following properties:
1) the $N$ single-particle states $\{\ket{\alpha_i\beta_i}\}$ are peaked in separated spatial regions $\{\mathcal{R}_i\}$ (see Figure \ref{Figure1});
2) $\langle \Psi^{(N)}|\Pi^{(N)}_{\alpha}|\Psi^{(N)}\rangle\neq 0$, i.e. the probability of obtaining the projected state must be different from zero (see appendix \ref{AppA}). 

We indicate with $P_{\alpha_i \chi_j}$ the single-particle probability that the result $\alpha_i$ comes from the state $\ket{\chi_j}$. We then define the joint probability $P_{\alpha\mathcal{P}}:=P_{\alpha_1\chi_{p_1}} P_{\alpha_2\chi_{p_2}}\cdots P_{\alpha_N\chi_{p_N}}$, where $\mathcal{P}=\{p_1,p_2,...,p_N\}$ is one of the $N!$ permutations of the $N$ single-particle states $\{\ket{\chi_i}\}$. Notice that $P_{\alpha\mathcal{P}}$ can be nonzero for each of the $N!$ permutations, since in general the outcome $\alpha_i$ can come from any of the single-particle state $\ket{\chi_j}$. The quantity $\mathcal{Z}=\sum_{\mathcal{P}}P_{\alpha\mathcal{P}}$ thus accounts for this no which-way effect concerning the probabilities. The degree of indistinguishability is finally given by 
\begin{equation}\label{indistinguishabilityN}
\mathcal{I}_{\alpha}:=-\sum_{\mathcal{P}}\dfrac{P_{\alpha \mathcal{P}}}{\mathcal{Z}^{(N)}}\log_2 \dfrac{P_{\alpha \mathcal{P}}}{\mathcal{Z}^{(N)}}.
\end{equation}
This quantity depends on measurements performed on the state. If the particles are initially all spatially separated, each in a different measurement region, only one permutation remains and $\mathcal{I}_{\alpha}=0$: we have complete knowledge on the single-particle state $\ket{\chi_j}$ which gives the outcome $\alpha_i$, meaning that the particles are distinguishable with respect to the measurement $\Pi_{\alpha}^{(N)}$. On the other hand, if for any possible permutation $\mathcal{P}'\neq \mathcal{P}$ one has $P_{\alpha \mathcal{P'}}=P_{\alpha \mathcal{P}}$, indistinguishability is maximum and reaches the value $\mathcal{I}_{\alpha}=\log_2N!$. As a specific example, when $\chi_i^a = \psi_i$ (spatial wave functions) and $\chi_i^b=\sigma_i$ (pseudospins), $\mathcal{I}_{\alpha}$ of Eq.~(\ref{indistinguishabilityN}) is the direct generalization of $\mathcal{I}_\mathrm{LR}$ of Eq.~(\ref{Indistinguishability}) and provides the degree of spatial indistinguishability under sLOCC for $N$ identical particles.

\textbf{Application: Noisy preparation of pure entangled state.} We now apply the tools above to a situation of experimental interest, namely noisy entanglement generation with identical particles. 

Werner state \cite{werner1989quantum} $W_\mathrm{AB}^{\pm}$ for two nonidentical qubits A and B is considered as the paradigmatic example of realistic noisy preparation of a pure entangled state subject to the action of white noise. In the usual formulation, it is defined as a mixture of a pure maximally entangled (Bell) state and of the maximally mixed state (white noise). Its explicit expression, assuming to be interested in generating the Bell state $\ket{\Psi_\pm^{\mathrm{AB}}}=(\ket{\uparrow_\mathrm{A},\downarrow_\mathrm{B}}\pm\ket{\downarrow_\mathrm{A},\uparrow_\mathrm{B}})/\sqrt{2}$, is 
$W_\mathrm{AB}^{\pm}=(1-p)\ket{\Psi_\pm^{\mathrm{AB}}}\bra{\Psi_\pm^{\mathrm{AB}}}+p\mathbb{I}_\mathrm{4}/4$, where $\mathbb{I}_\mathrm{4}$ is the $4\times4$ identity matrix and $p$ is the noise probability which accounts for the amount of white noise in the system during the pure state preparation stage. The Werner state $W_\mathrm{AB}^{\pm}$ is also the product of a single-particle depolarizing channel induced by the environment applied to an initial Bell state \cite{audretsch,horo2009}. It is known that the concurrence for such state is  $C(W_\mathrm{AB}^{\pm})=1-3p/2$ when $0\leq p< 2/3$, being zero otherwise \cite{horo2009} (see black dot-dashed line of 
Fig.~\ref{fig:figure3}\textbf{a}).

In perfect analogy, the Werner state for two identical qubits with spatial wave functions $\psi_1$, $\psi_2$ can be defined by
\begin{gather}\label{Wind}
   \mathcal{W}^{\pm}=\left(1-p\right)\ket{1_{\pm}}\bra{1_{\pm}}+p\mathcal{I}_\mathrm{4}/4,
\end{gather}
where $\mathcal{I}_\mathrm{4}=\sum_{i=1,2; s=\pm} \ket{i_s}\bra{i_s}$, having used the orthogonal Bell-state basis $\mathcal{B}_\mathrm{\{ 1_{\pm},2_{\pm}\}}=\{\ket{1_+}, \ket{1_-}, \ket{2_+}, \ket{2_-}\}$
with
\begin{gather}
        \ket{1_{\pm}}:=(\ket{\psi_1\uparrow,\psi_2\downarrow}\pm\ket{\psi_1\downarrow,\psi_2\uparrow})/\sqrt{2},\nonumber
    \\ 
        \ket{2_{\pm}}:=(\ket{\psi_1\uparrow,\psi_2\uparrow})\pm
        \ket{\psi_1\downarrow,\psi_2\downarrow})/\sqrt{2}.
\label{generalizedBS}
\end{gather}
The Werner state of Eq.~(\ref{Wind}) is justified as a model of noisy state. In fact, it is straightforward to see that $\mathcal{W}^{\pm}$ is produced by a localized depolarizing channel acting on one of two initially separated identical qubits, followed by a quick single-particle spatial deformation procedure which makes the two identical qubits spatially overlap (see appendix \ref{AppB}). Hence, in Eq.~(\ref{Wind}), $\ket{1_{\pm}}$ is the target pure state to be prepared and $\mathcal{I}_4/4$ is the noise as a mixture of the four Bell states. 

\begin{figure}[!t]
\begin{center}
\includegraphics[width=0.435\textwidth]{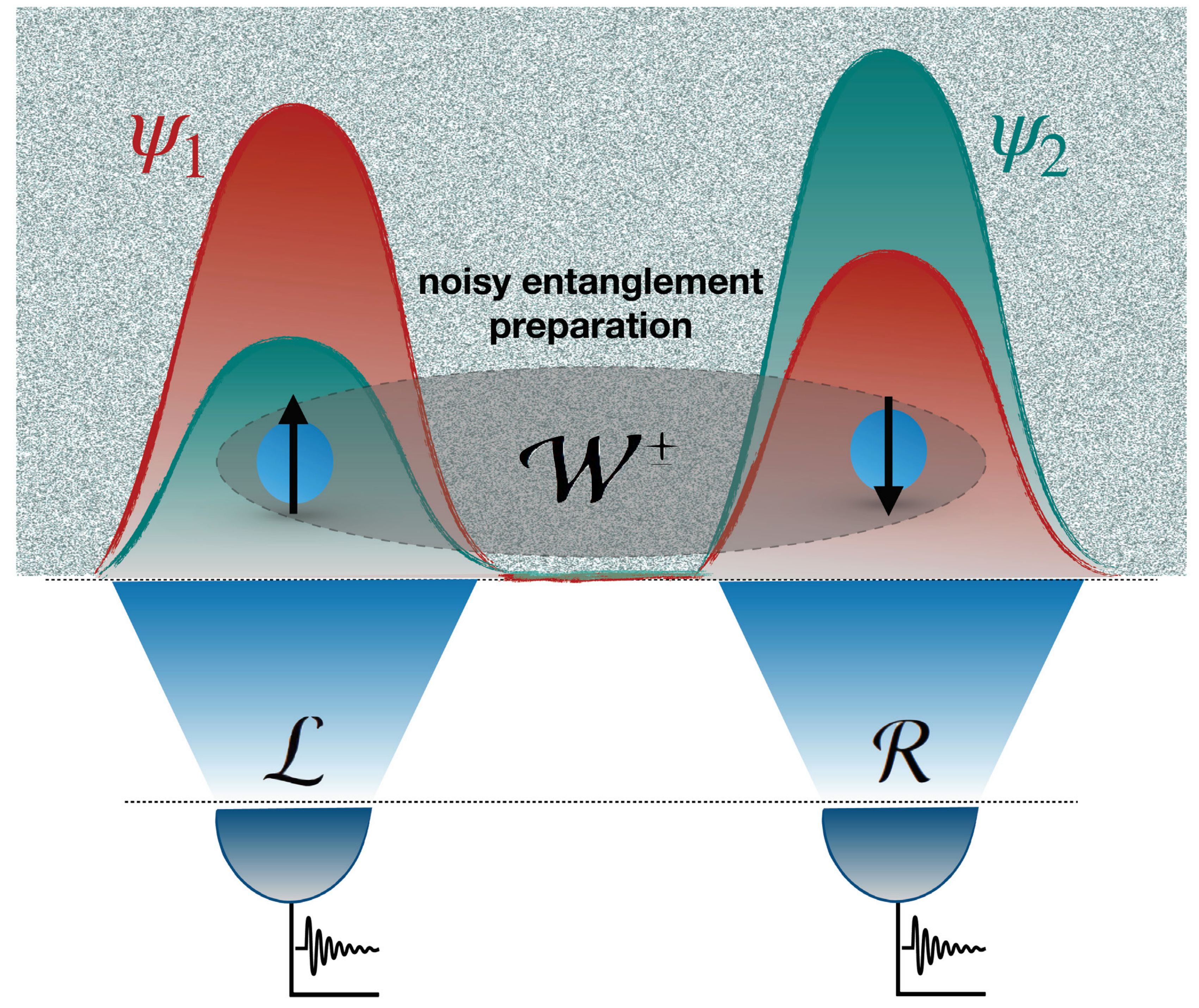}
\caption{\textbf{Noisy entanglement preparation with tailored spatial indistinguishability.} Illustration of two controllable spatially overlapping wave functions $\psi_1$, $\psi_2$ peaked in the two localized regions of measurement 
$\mathcal{L}$ and $\mathcal{R}$. The two identical qubits are prepared in an entangled state under noisy conditions, giving $\mathcal{W}^{\pm}$. The degree of spatial indistinguishability can be tuned, being $0\leq\mathcal{I}_\mathrm{LR}\leq 1$.}
\label{Figure2}
\end{center}
\end{figure}

Given the configuration of the spatial wave functions and using the sLOCC framework, the amount of operational entanglement contained in $\mathcal{W}^{\pm}$ can be obtained by the concurrence $C_\mathrm{LR}(\mathcal{W}^{\pm})=C(\mathcal{W}_\mathrm{LR}^{\pm})$ of Eq. \eqref{CLR}. 
Notice that the state of Eq. \eqref{Wind} is in general not normalized, depending on the specific spatial degrees of freedom \cite{compagnoRSA}. 
This is irrelevant at this stage, since the entanglement of $\mathcal{W}^{\pm}$ is calculated on the final distributed state $\mathcal{W}^{\pm}_\mathrm{LR}$, which is obtained from $\mathcal{W}^{\pm}$ after sLOCC and is normalized (see Eq. \eqref{rhoLR}). 
Focusing on the observation of entanglement, a well-suited configuration for the spatial wave functions is $\ket{\psi_1}=l\ket{\mathrm{L}}+r\ket{\mathrm{R}}$ and $\ket{\psi_2}=l'\ket{\mathrm{L}}+r'e^{i\theta}\ket{\mathrm{R}}$, where $l$, $r$, $l'$, $r'$ are non-negative real numbers ($l^2+r^2=l'^2+r'^2=1$) and $\theta$ is a phase. The wave functions are thus peaked in the two localized measurement regions $\mathcal{L}$ and $\mathcal{R}$, as depicted in Figure \ref{Figure2}. The degree of spatial indistinguishability $\mathcal{I}_\mathrm{LR}$ of Eq. \eqref{Indistinguishability} is tailored by adjusting the shapes of $\ket{\psi_1}$, $\ket{\psi_2}$, with $P_{\mathrm{L}\psi_1}=l^2$, $P_{\mathrm{L}\psi_2}=l'^2$ (implying $P_{\mathrm{R}\psi_1}=r^2$, $P_{\mathrm{R}\psi_2}=r'^2$). The interplay between $C_\mathrm{LR}(\mathcal{W}^{\pm})$ and $\mathcal{I}_\mathrm{LR}$ versus noise probability $p$ can be then investigated (some explicit expressions of $C_\mathrm{LR}(\mathcal{W}^{\pm})$ are reported in appendix \ref{AppC}).

\begin{figure*}[t!] 
\includegraphics[width=0.9\textwidth]{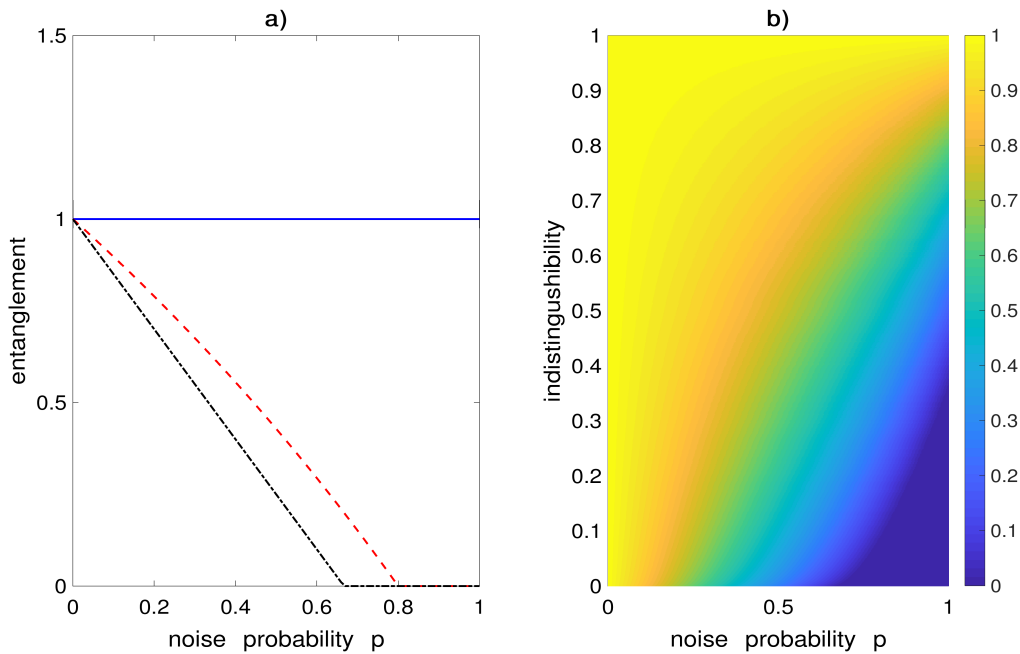}
\caption{\textbf{Prepared entanglement and indistinguishability.} \textbf{a.} Entanglement $C_\mathrm{LR}(\mathcal{W}^{\pm})$ as a function of noise probability $p$ for different degrees of spatial indistinguishability $\mathcal{I}_\mathrm{LR}$ and system parameters: blue solid line is for target state $\ket{1_{-}}$, $\mathcal{I}_\mathrm{LR}=1$ ($l=l'$), fermions (with $\theta=0$) or bosons (with $\theta=\pi$); red dashed line is for target state $\ket{1_{+}}$, $\mathcal{I}_\mathrm{LR}=1$ ($l=l'$), fermions (with $\theta=\pi$) or bosons (with $\theta=0$); black dot-dashed line is for distinguishable qubits ($\mathcal{I}_\mathrm{LR}=0$, $l=1$ and $l'=0$ or vice versa). \textbf{b.} Contour plot of entanglement $C_\mathrm{LR}(\mathcal{W}^{-})$ versus noise probability $p$ and spatial indistinguishability $\mathcal{I}_\mathrm{LR}$ for target state $\ket{1_{-}}$, fermions (with $\theta=0$) or bosons (with $\theta=\pi$), fixing $l=r'$.}
\label{fig:figure3}
\end{figure*}

Generally, the entanglement amount is conditional since the state is obtained by postselection. As a result, the entangled state $\rho_{LR}$ is detectable if the sLOCC probability $P_\mathrm{LR}$ is high enough to be of experimental relevance. Let us see what happens for $\mathcal{I}_\mathrm{LR}=1$ ($l=l'$). 
When the target pure state in Eq. \eqref{Wind} is $\ket{1_{-}}$, using in Eq. \eqref{generalizedBS} the explicit expressions of $\ket{\psi_1}$, $\ket{\psi_2}$ with $\theta=0$ ($\theta=\pi$) for fermions (bosons), from $\mathcal{W}^{-}$ we obtain by sLOCC  the distributed Bell state $\mathcal{W}^{-}_\mathrm{LR}=\ket{1_{-}^\mathrm{LR}}\bra{1_{-}^\mathrm{LR}}$, with $\ket{1_{-}^\mathrm{LR}}=(\ket{\mathrm{L}\uparrow,\mathrm{R}\downarrow}-\ket{\mathrm{L}\downarrow,\mathrm{R}\uparrow})/\sqrt{2}$, therefore (see blue solid line of Fig. \ref{fig:figure3}\textbf{a})
\begin{equation} \label{eq:10}
  C_\mathrm{LR}(\mathcal{W}^{-})=C(\mathcal{W}^{-}_\mathrm{LR})=1,\ \textit{for any noise probability}\ p
\end{equation}
for which the probabilities of detecting this state for fermions and bosons are, respectively,
\begin{equation}
    P_\mathrm{LR}^{\mathrm{(f)}}=2l^2(1-l^2),\quad
    P_\mathrm{LR}^{\mathrm{(b)}}=\frac{2 l^2 \left(1-l^2\right) (4-3 p)}{2-\left(1-2l^2\right)^2 (2-3 p)}.
\end{equation}
Notice that the sLOCC probability for fermions, $P_\mathrm{LR}^{\mathrm{(f)}}$, is in this case independent of the noise probability. Fixing $l^2=1/2$ we maximize the sLOCC probability, which is $1/2$ for fermions and $1/4$ for bosons in the worst scenario of maximum noise probability $p=1$. Differently, targeting the pure state $\ket{1_{+}}$ in 
Eq.~(\ref{Wind}), for fermions (bosons) with $\theta=\pi$ ($\theta=0$), one gets a $p$-dependent $\mathcal{W}^{+}_{\mathrm{LR}}$ by sLOCC with $C_\mathrm{LR}(\mathcal{W}^{+})=C(\mathcal{W}^{+}_\mathrm{LR})=(4-5p)/(4-p)$ when $0 \leq p < 4/5$, being zero elsewhere. The entanglement now decreases with increasing noise, remaining however larger than that for nonidentical qubits (see red dashed line of 
Fig.~\ref{fig:figure3}\textbf{a}). The choice of the state to generate makes a difference concerning noise protection by indistinguishability. However, we remark that identical qubits in the distributed resource state after sLOCC, $\mathcal{W}^{\pm}_\mathrm{LR}$, are individually addressable. Local unitary operations (rotations) in $\mathcal{L}$ and $\mathcal{R}$ can be applied to each qubit to transform the noise-free prepared $\ket{1_{-}^\mathrm{LR}}$ into any other Bell state \cite{horo2009}. Another relevant aspect is that the phase $\theta$ in $\ket{\psi_2}$ acts as a switch between fermionic and bosonic behavior of entanglement (see appendix \ref{AppC} for details on more general instances). The result for nonidentical particles is retrieved when the qubits become distinguishable ($\mathcal{I}_{\mathrm{LR}}=0$, $l=r'=1$ or $l=r'=0$).

Since the preparation of $\ket{1_{-}}$, as represented by Eq.~(\ref{Wind}), results to be noise-free for both fermions and bosons when $\mathcal{I}_\mathrm{LR}=1$, it is important to know what occurs for a realistic imperfect degree of spatial indistinguishability. In Fig.\ref{fig:figure3}\textbf{b} we display entanglement as a function of both $p$ and $\mathcal{I}_\mathrm{LR}$. The plot reveals that entanglement preparation can be efficiently protected against noise also for $\mathcal{I}_\mathrm{LR}<1$. 
A crucial information in this scenario is the minimum degree of $\mathcal{I}_\mathrm{LR}$ that guarantees nonlocal entanglement in $\mathcal{L}$ and $\mathcal{R}$, by violating a CHSH-Bell inequality \cite{horo2009}, whatever the noise probability $p$. We remark that a Bell inequality violation based on sLOCC provides a faithful test of local realism \cite{sciarrinoPRA}.
Using the Horodecki criterion \cite{horodecki1995violating}, we find that the Bell inequality is violated for any $p$ whenever $0.76< \mathcal{I}_\mathrm{LR} \leq 1 $, implying $0.56< C_\mathrm{LR}(\mathcal{W}^{-}) \leq 1 $ (see appendix \ref{AppD} for details).
This is basically different from the case of distinguishable qubits where, as known \cite{horo2009}, $W_\mathrm{AB}^{\pm}$ violates Bell inequality only for small white noise probabilities $0 \leq p< 0.292$ (giving $0.68 < C(W_\mathrm{AB}^{\pm}) \leq 1$). 
These results show robust quantum entanglement preparation against noise through spatial indistinguishability, even partial. In fact, rather than addressing individual qubits, here one controls the shapes of their spatial wave functions $\ket{\psi_1}$, 
$\ket{\psi_2}$. Significant changes in these shapes can occur that anyway maintain $\mathcal{I}_\mathrm{LR}$ of Eq.~(\ref{Indistinguishability}) beyond the threshold ($\approx 0.76$) assuring noise-free generation of nonlocal entanglement. Indistinguishability here emerges as a property of composite quantum systems inherently robust to surrounding-induced disorder, protecting exploitable quantum correlations.

\section{Discussion}

In this work, we have studied the effect of spatial indistinguishability on entanglement preparation under noise, within the sLOCC framework. Firstly, thanks to the analogy with known methods for distinguishable particles, the entanglement of formation, and the related concurrence, has been defined for an arbitrary pure or mixed state of two identical qubits. Secondly, we have introduced the degree of spatial indistinguishability of identical particles by an entropic measure of information. This achievement entails a continuous quantitative identification of indistinguishability as an informational resource. 
Hence, one can evaluate the amount of entanglement exploitable by sLOCC into two separated operational sites under general conditions of spatial indistinguishability and state mixedness. 

The Werner state $\mathcal{W}^{\pm}$ has been then chosen as a typical instance of noisy mixed state of two identical qubits, with tunable spatial overlap of their wave functions on the two remote operational regions. 
The tunable spatial overlap rules the indistinguishability degree. We have found that, under conditions of complete spatial indistinguishability, maximally entangled pure states between internal (spin-like) degrees of freedom can be prepared unaffected by noise. Even in the more realistic scenario of experimental errors in controlling particle spatial overlap, we have supplied a lower bound for the degree of spatial indistinguishability beyond which the generated entangled state violates the CHSH-Bell inequality independently of the amount of noise. These findings are independent of particle statistics, holding for both bosons and fermions.
One reasonably may expect that also coherence can be protected by spatial indistinguishability, based on a previous work showing that the latter enables quantum coherence \cite{castellini2019indistinguishability}. This supports the observed effects in an experiment of coherence endurance due to particle indistinguishability \cite{Perez-Leija2018}.

The degree of spatial indistinguishability exhibits robustness to variations in the configuration of spatial wave functions, being then capable of shielding nonlocal entangled states against preparation noise. 
Therefore, indistinguishability represents a resource of quantum networks made of identical qubits enabling noise-free entanglement generation by its physical nature. Such a finding, which is promising to fault-tolerant quantum information tasks under environmental noise, adds to other known protection techniques of quantum states based on, for example, topological properties \cite{KITAEV20032,PhysRevLett.98.160502,Nigg302,PhysRevLett.99.020503,gladchenko2009superconducting,mittal2018topological,
wang2019topologically,science2018}, dynamical decoupling or decoherence-free subspace \cite{lidarReview,viola1999dynamical,zanardiPRL,lidar1998decoherence,
lofrancoPRB}. 
As an outlook, the effects of spatial indistinguishability on quantumness protection for different types of environmental noises will be addressed elsewhere.

Various experimental contexts can be thought for implementing the above theoretical scenario.
For example, in quantum optics, spatially localized detectors can perform the required measurements while beam splitters can serve as controller of spatial wave functions of independent traveling photons (bosons) with given polarization pseudospin. In a more sophisticated example with circular polarizations, one may employ orbital angular momentum of photons as spatial wave function and spin angular momentum as spin-like degree of freedom \cite{mair2001entanglement, nagali2009quantum,aolita2007quantum}. Setups using integrated quantum optics can also simulate fermionic statistics using photons \cite{sansoni2012two}. Other suitable platforms for fermionic subsystems can be supplied either by superconducting quantum circuits with Ramsey interferometry \cite{PhysRevLett.115.260403}, or by quantum electronics with quantum point contacts as electronic beam splitters \cite{bocquillon2013coherence,PhysRevLett.121.166801,electronsReview}. 
The results of this work are expected to stimulate further theoretical and experimental studies concerning the multiple facets of indistinguishability as a controllable fundamental quantum trait and its exploitation for quantum technologies.


\appendix

\section{Amplitudes and probabilities in the no-label approach}\label{AppA}

For calculating all the necessary probabilities and traces to obtain the results of the work, under different spatial configurations of the wave functions, we need to compute scalar products (amplitudes) between states of $N$ identical particles. 

The $N$-particle probability amplitude has been defined in the literature by means of the no-label particle-based approach, here adopted, to deal with systems of identical particles \cite{compagnoRSA}. Indicating with $\chi_i$, $\chi'_i$ ($i=1,\ldots, N$) single-particle states containing all the degrees of freedom of the particle, the general expression of the $N$-particle probability amplitude is
\begin{eqnarray}\label{Nampleta}
&\langle \chi'_1,\chi'_2,\ldots,\chi'_n|\chi_1,\chi_2,\ldots,\chi_n\rangle & \nonumber\\ & :=\sum_P\eta^P\langle \chi'_1|\chi_{P_1}\rangle\langle \chi'_2|\chi_{P_2}\rangle \ldots \langle \chi'_n|\chi_{P_n}\rangle, &
\end{eqnarray}
where $P=\{P_1,P_2,...,P_n\}$ in the sum runs over all the one-particle state permutations, $\eta=\pm1$ for bosons and fermions, respectively, and $\eta^P$ is 1 for bosons and 1 (-1) for even (odd) permutations for fermions. Notice that the explicit dependence on the particle statistics appears, as expected. 

Along our manuscript, we especially need two-particle probabilities and trace. For $N=2$, the general expression above reduces to the following two-particle probability amplitude
\begin{equation}\label{2ampleta}
\langle \chi'_1,\chi'_2|\chi_1,\chi_2\rangle =\langle \chi'_1|\chi_{1}\rangle\langle \chi'_2|\chi_{2}\rangle+\eta \langle \chi'_1|\chi_{2}\rangle\langle \chi'_2|\chi_{1}\rangle. 
\end{equation}

\section{Werner state $\mathcal{W}^\pm$ of two indistinguishable qubits}\label{AppB}

In the following, we describe two different ways which produce the Werner state $\mathcal{W}^\pm$ of two indistinguishable particles, given in the main text.

\textbf{Analogy with distinguishable particles.}
Let us consider the orthogonal Bell states of two indistinguishable qubits, with spatial wave functions $\psi_1$ and $\psi_2$, which are defined as
\begin{eqnarray}
\ket{1_{\pm}}&=&(\ket{\psi_1\uparrow,\psi_2\downarrow}\pm
\ket{\psi_1\downarrow,\psi_2\uparrow})/\sqrt{2},\nonumber\\
\ket{2_{\pm}}&=&(\ket{\psi_1\uparrow,\psi_2\uparrow})\pm
\ket{\psi_1\downarrow,\psi_2\downarrow})/\sqrt{2}.
\label{BS}
\end{eqnarray}
Each of these Bell states is not normalized in general. Their normalized expressions are
\begin{eqnarray}
\ket{\bar{1}_{\pm}}&=&\dfrac{1}{\sqrt{2\mathcal{N}_{1_{\pm}}}}
(\ket{\psi_1\uparrow,\psi_2\downarrow}\pm\ket{\psi_1\downarrow,\psi_2\uparrow}),\nonumber\\
\ket{\bar{2}_{\pm}}&=&\dfrac{1}{\sqrt{2\mathcal{N}_{2_{\pm}}}}
(\ket{\psi_1\uparrow,\psi_2\uparrow})\pm\ket{\psi_1 \downarrow,\psi_2\downarrow}),
\label{generalizedBS}
\end{eqnarray}
where $\mathcal{N}_{1_-}=(1-\eta |\langle \psi_1|\psi_2\rangle|^2)$ and 
$\mathcal{N}_{1_+}=\mathcal{N}_{2_{\pm}}=
(1+\eta |\langle \psi_1|\psi_2\rangle|^2)$.

For two distinguishable (or nonidentical) qubits A and B, a Werner state $W_\mathrm{AB}^\pm$ is a mixture of a pure maximally entangled (Bell) state and of the maximally mixed state \cite{werner1989quantum}. Its explicit expression, assuming to be interested in preparing the Bell state $\ket{\Psi_\pm^{\mathrm{AB}}}=(\ket{\uparrow_\mathrm{A},\downarrow_\mathrm{B}}\pm\ket{\downarrow_\mathrm{A},\uparrow_\mathrm{B}})/\sqrt{2}$, is 
\begin{equation}\label{WAB}
W_\mathrm{AB}^\pm=(1-p)\ket{\Psi_\pm^{\mathrm{AB}}}\bra{\Psi_\pm^{\mathrm{AB}}}+p\mathbb{I}_\mathrm{4}/4,
\end{equation}
where $\mathbb{I}_\mathrm{4}$ is the $4\times4$ identity matrix and $p$ is the noise probability accounting for the amount of white noise in the system \cite{horo2009}. Notice that $\mathbb{I}_\mathrm{4}$ can be written either in the computation basis or in the basis of the four Bell states. The Werner state is of wide interest since it can be meant as a state representing a (realistic) noisy preparation of pure two-qubit entangled states \cite{audretsch,horo2009}.

In strict analogy with $W_\mathrm{AB}^\pm$, the Werner state of a pair of identical qubits with spatial wave functions $\psi_1$ and $\psi_2$ can be expressed in the orthogonal Bell-state basis $\mathcal{B}_\mathrm{\{ 1_{\pm},2_{\pm}\}}=\{\ket{1_+}, \ket{1_-}, \ket{2_+}, \ket{2_-}\}$ by
\begin{equation}\label{WindSI}
   \mathcal{W}^\pm=(1-p)\ket{1_{\pm}}\bra{1_{\pm}}+p\mathcal{I}_\mathrm{4}/4,
\end{equation}
where $\mathcal{I}_\mathrm{4}=\sum_{i=1,2; s=\pm} \ket{i_s}\bra{i_s}$. The state $\mathcal{W}$ is in general unnormalized. Its normalized expression $\bar{\mathcal{W}}=\mathcal{W}/\mathcal{N}_\pm$ requires a global normalization constant  
\begin{gather}
  \mathcal{N}_\pm=1+\eta |\langle \psi_1|\psi_2\rangle|^2[p/2\pm(1-p)].
\end{gather}
Using normalized Bell states of Eq. \eqref{generalizedBS}, this state can be equivalently expressed by
\begin{gather}\label{WindNorm}
\bar{\mathcal{W}^\pm}=\dfrac{1}{\mathcal{N}_\pm}\left((1-p)\mathcal{N}_{1_\pm}\ket{\bar{1}_\pm}\bra{\bar{1}_\pm}+\dfrac{p}{4}\sum_{i=1,2;s=\pm} \mathcal{N}_{i_s} \ket{\bar{i}_s}\bra{\bar{i}_s}\right).
\end{gather}
Notice that the state $\mathcal{W}^\pm$ reduces to the usual Werner state of two distinguishable qubits $W_\mathrm{AB}^\pm$ of Eq. \eqref{WAB} when there is no spatial overlap between the qubits, so that the latter can be individually addressed in their separated locations.

\textbf{$\mathcal{W}^\pm$ as a model of noisy state.}
The Werner state of Eq.~(\ref{WindSI}) is justified as a model of noisy state, being produced by a localized single-particle depolarizing channel acting on one of two initially separated identical qubits, followed by a quick single-particle spatial deformation procedure which makes the two identical qubits spatially overlap. In the following we show this in detail.

It is known that the Werner state $W_{AB}^{\pm}=(1-p)\ket{\Psi_\pm^{AB}}\bra{\Psi_\pm^{AB}}+p\mathbb{I}_4/4$ for two distinguishable (nonidentical) qubits $A$ and $B$ is the output of the single-qubit depolarizing channel action on one of the two qubits of an initial Bell state \cite{audretsch}. The process is such that the initial (maximally entangled) pure state tends to remain in the initial state with probability $1-p$ and changes to the maximally mixed state $\mathbb{I}_4/4$ with probability $p$. 
The Kraus (operator-sum) representation of the single-qubit depolarizing channel acting on an elementary pure state like $\rho_{AB}=\ket{\phi_A,\phi_B}\bra{\phi_A,\phi_B}$ is given by \cite{chuang2010,audretsch}
\begin{equation}\label{Kraus}
  \rho(t)=\sum_{i=0}^4\ket{K^{A}_i\phi_A}\otimes\ket{\mathbb{I}_2^{B}\phi_B}\bra{K^{A}_i\phi_A}\otimes\bra{\mathbb{I}_2^{B}\phi_B},
\end{equation}  
where the Kraus operators are $K_0=\sqrt{1-3p/4}\ \mathbb{I}_2$ and $K_i=\sqrt{p/4}\sigma_i$ ($i=1,2,3$), with $\mathbb{I}_2$ being the single-particle ($2\times2$) identity operator and $\sigma_i$ the usual Pauli matrices. Notice that here the channel individually acts on the internal degrees of freedom (pseudospins) of qubit $A$ and $\bra{K^{A}_i\phi_A}=\bra{\phi_A}K_i^{A\dagger}$. Using the action on the elementary state of Eq.~(\ref{Kraus}), it is then straightforward to see that $W_{AB}^{\pm}$ is obtained by the single-qubit depolarizing channel starting from a Bell state $\ket{\Psi_\pm^{AB}}=(\ket{\uparrow_\mathrm{A},\downarrow_\mathrm{B}}\pm\ket{\downarrow_\mathrm{A},\uparrow_\mathrm{B}})/\sqrt{2}$ \cite{audretsch}. 

Now, let us consider a pair of two identical qubits which are initially separated, for example one in a location $\mathrm{L}_1$ and one a in location $\mathrm{L}_2$. If these qubits are initially in a Bell state $\ket{\Psi_\pm^{\mathrm{L_1L_2}}}=(\ket{\mathrm{L}_1\uparrow,\mathrm{L}_2\downarrow}\pm\ket{\mathrm{L}_1\downarrow,\mathrm{L}_2\uparrow})/\sqrt{2}$ and the localized single-particle depolarizing channel of Eq.~(\ref{Kraus}) acts on $\mathrm{L}_1$ ($A\leftrightarrow\mathrm{L}_1$, $B\leftrightarrow\mathrm{L}_2$), for instance, the qubits behave as distinguishable particles labeled, respectively, by $\mathrm{L}_1$ and $\mathrm{L}_2$. Therefore, the resulting state at this stage is $W_\mathrm{L_1L_2}^{\pm}=(1-p)\ket{\Psi_\pm^{\mathrm{L_1L_2}}}\bra{\Psi_\pm^{\mathrm{L_1L_2}}}+p\mathbb{I}_{4}/4$, where $\mathbb{I}_{4}=\sum_{k=\pm}(\ket{\Psi_k^{\mathrm{L_1L_2}}}\bra{\Psi_k^{\mathrm{L_1L_2}}}
+\ket{\Phi_k^{\mathrm{L_1L_2}}}\bra{\Phi_k^{\mathrm{L_1L_2}}})$, with $\ket{\Phi_k^{\mathrm{L_1L_2}}}=(\ket{\mathrm{L}_1\uparrow,\mathrm{L}_2\uparrow}\pm\ket{\mathrm{L}_1\downarrow,\mathrm{L}_2\downarrow})/\sqrt{2}$ being the other couple of Bell states. After the action of this localized depolarizing channel, one can apply a quick spatial deformation procedure on each qubit making the following transformation on the spatial degree of freedom: $\ket{\mathrm{L_1}}\rightarrow \ket{\psi_1}$, $\ket{\mathrm{L_2}}\rightarrow \ket{\psi_2}$. The two final spatial wave functions $\ket{\psi_1}$, $\ket{\psi_2}$ are in general spatially overlapping and the final output state of the entire procedure (localized depolarizing channel plus spatial deformation) immediately results to be 
\begin{equation}\label{WindSI2}
   \mathcal{W}^\pm=(1-p)\ket{1_{\pm}}\bra{1_{\pm}}+p\mathcal{I}_\mathrm{4}/4,
\end{equation}
which is just the Werner state defined in Eq.~(\ref{WindSI}). This result justifies the introduced Werner state for indistinguishable particles, $\mathcal{W}^\pm$, as a proper noisy state. We stress that a general analysis of different environmental noises and more general derivations of noise models for systems of indistinguishable particles is beyond the aim of the present paper and will be provided elsewhere.

\section{Concurrence of $\mathcal{W}^\pm$ state}\label{AppC}

Given the state $\mathcal{W}^\pm$ of Eq. \eqref{WindSI} and the structure of the spatial wave functions, the amount of operational entanglement contained in $\mathcal{W}^\pm$ can be obtained by sLOCC. Aiming at observing entanglement in separated regions, also in view of CHSH-Bell inequality violations \cite{horo2009}, a suitable form for the spatial wave functions appearing in Eq. \eqref{WindSI} is 
\begin{equation}\label{spatialWF}
\ket{\psi_1}=l\ket{\mathrm{L}}+r\ket{\mathrm{R}},\quad
\ket{\psi_2}=l'\ket{\mathrm{L}}+r'e^{i\theta}\ket{\mathrm{R}},
\end{equation}
where $l$, $r$, $l'$ and $r'$ are non-negative real numbers (with $l^2+r^2=l'^2+r'^2=1$) and $\theta$ is a phase. 

\begin{figure*}[t!] 
\includegraphics[width=0.88\textwidth]{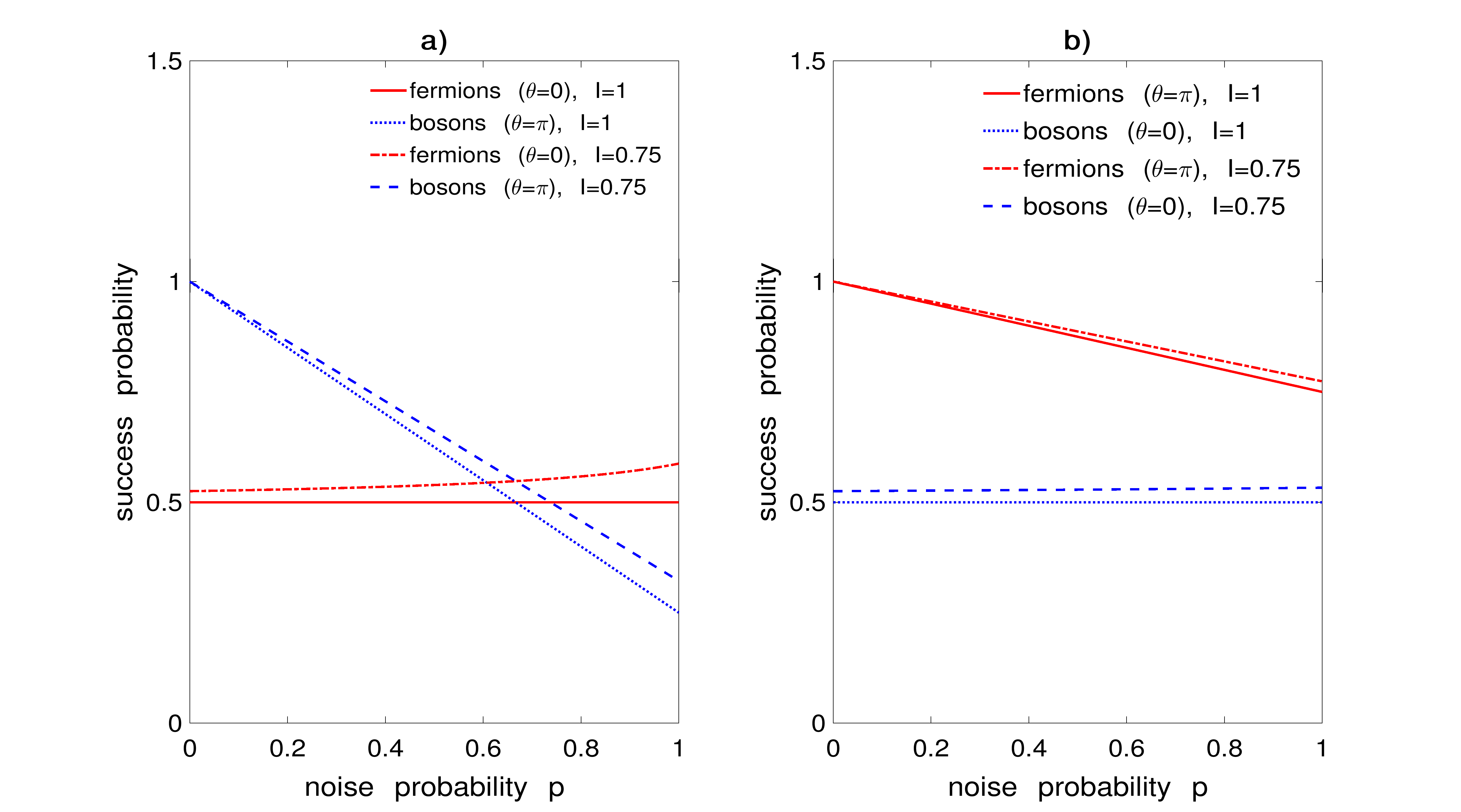}
\caption{sLOCC probability $P_\mathrm{LR}$ (success probability) as a function of noise probability $p$ for some degrees of spatial indistinguishability $\mathcal{I}_\mathrm{LR}$ and particle statistics, fixing $l=r'$ in the wave functions of Eq. \eqref{spatialWF}. \textbf{a.} Target pure state $\ket{1_{-}}$ in Eq. \eqref{WindSI}. \textbf{b.} Target pure state $\ket{1_{+}}$ in Eq. \eqref{WindSI}.}
\label{fig:Pr}
\end{figure*}

Projecting $\mathcal{W}$ onto the (operational) subspace spanned by the computational basis $\mathcal{B}_\mathrm{LR}=\{\ket{\mathrm{L}\uparrow,\mathrm{R}\uparrow}, \ket{\mathrm{L}\uparrow,\mathrm{R}\downarrow}, \ket{\mathrm{L}\downarrow,\mathrm{R}\uparrow}, \ket{\mathrm{L}\downarrow,\mathrm{R}\downarrow}\}$ by means of the projector
\begin{equation}\label{PiLRSI}
\Pi_{\mathrm{LR}}^{(2)}=\sum_{\tau_1,\tau_2=\uparrow,\downarrow}\ket{\mathrm{L}\tau_1,\mathrm{R}\tau_2}\bra{\mathrm{L}\tau_1,\mathrm{R}\tau_2},
\end{equation}
and using the scalar product of Eq. \eqref{2ampleta}, we obtain the distributed resource state
\begin{equation}\label{WLR}
\mathcal{W}^\pm_\mathrm{LR}=\Pi_\mathrm{LR}^{(2)}\mathcal{W}^\pm\Pi_\mathrm{LR}^{(2)}/\mathrm{Tr}(\Pi_\mathrm{LR}^{(2)}\mathcal{W}^\pm), 
\end{equation}
with probability $P_\mathrm{LR}=\mathrm{Tr}(\Pi_\mathrm{LR}^{(2)}\bar{\mathcal{W}^\pm})$. 
The trace operation is performed in the LR-subspace. The (normalized) state $\mathcal{W}^\pm_\mathrm{LR}$ can be then treated as the state of two distinguishable qubits in separated bound states $\ket{\mathrm{L}}$, localized in the region $\mathcal{L}$, and $\ket{\mathrm{R}}$, localized in the region $\mathcal{R}$ \cite{PhysRevLett.120.240403}. 
Therefore, the sLOCC-based concurrence $C_\mathrm{LR}(\mathcal{W}^\pm):=C(\mathcal{W}^\pm_\mathrm{LR})$ can be calculated by the usual criterion for distinguishable qubits \cite{PhysRevLett.78.5022,PhysRevLett.80.2245}, that is by
\begin{equation}
C(\mathcal{W}^\pm_\mathrm{LR})=\max \{0,\sqrt{\lambda_4}-\sqrt{\lambda_3}-\sqrt{\lambda_2}-\sqrt{\lambda_1}\},
\end{equation}
where the $\lambda_i$'s are the eigenvalues, in decreasing order, of the non-Hermitian matrix $R=\rho_\mathrm{LR}\Tilde{\rho}_\mathrm{LR}$, being 
 \begin{equation}
     \Tilde{\rho}_\mathrm{LR}=\sigma^\mathrm{L}_{y}\otimes\sigma_{y}^\mathrm{R} 
\rho_\mathrm{LR}^{*}\sigma_{y}^\mathrm{L}\otimes\sigma_{y}^\mathrm{R},
\end{equation}
with localized Pauli matrices defined as $\sigma_{y}^\mathrm{L}\equiv\ket{\mathrm{L}}\bra{\mathrm{L}}\otimes\sigma_{y}$ and $\sigma_{y}^\mathrm{R}\equiv\ket{\mathrm{R}}\bra{\mathrm{R}}\otimes\sigma_{y}$. 

Thanks to the explicit expressions of the spatial wave functions of Eq. \eqref{spatialWF} above, the degree of spatial indistinguishability under sLOCC $\mathcal{I}_\mathrm{LR}$, defined in Eq. (9) of the main text, can be then explicitly calculated by substituting $P_{\mathrm{L}\psi_1}=l^2$, $P_{\mathrm{L}\psi_2}=l'^2$, $P_{\mathrm{R}\psi_1}=r^2$ and $P_{\mathrm{R}\psi_2}=r'^2$.

We can now report the explicit expressions of $C_\mathrm{LR}(\mathcal{W}^\pm):=C(\mathcal{W}^\pm_\mathrm{LR})$ for some cases of particular interest. 
As a remarkable aspect, our calculations show that the phase $\theta$ of $\ket{\psi_2}$, when assuming binary values $0,\pi$, acts as a switch between fermionic and bosonic entanglement behavior. This is why in the main text and in the following expressions, only these values of $\theta$ are chosen.

When the pure state $\ket{1_-}$ is considered in the $\mathcal{W}^{-}$ state of Eq. \eqref{WindSI}, for fermions (bosons) with $\theta=0$ ($\theta=\pi$) we find
\begin{gather}\label{conc1}
C(\mathcal{W}^{-}_\mathrm{LR})=\max\left\{0,\frac{(4-3 p)(lr'+l'r)^2-3p(lr'-l'r)^2}{4\left(l^2 r'^2+l'^2r^2+lr'rl'(2-3p)\right)}\right\},
\end{gather}
with probability
\begin{equation}\label{ProbLR1minus}
    P_\mathrm{LR}=\frac{2\left(l^2 r'^2+l'^2r^2+lr'rl'(2-3p)\right)}{2-\eta(2-3p)(ll'-\eta rr')^2},
\end{equation}
where $\eta=+1$ for bosons and $\eta=-1$ for fermions, as said before. 
The plot of this concurrence as a function of $\mathcal{I}_{LR}$ and $p$, fixing $l=r'$, is reported in Figure 3\text{b} of the main text.    
As can be seen from Eq. \eqref{conc1}, by setting $l=l'$ (implying $r=r'$) so to have maximum indistinguishability $\mathcal{I}_\mathrm{LR}=1$, one gets $C(\mathcal{W}^{-}_\mathrm{LR})=1$ independently of $p$. The sLOCC probability is maximized when $l=l'=1/\sqrt{2}$, taking the values $P_\mathrm{LR}=1/2$ for fermions and $P_\mathrm{LR}=1-3p/4$ for bosons. 
The sLOCC probability $P_\mathrm{LR}$ of Eq. \eqref{ProbLR1minus} is plotted in Fig. \ref{fig:Pr}(a) as a function of noise probability $p$ for some degrees of spatial indistinguishability $\mathcal{I}_\mathrm{LR}$. 

\begin{figure}[t!] 
\includegraphics[width=0.48\textwidth]{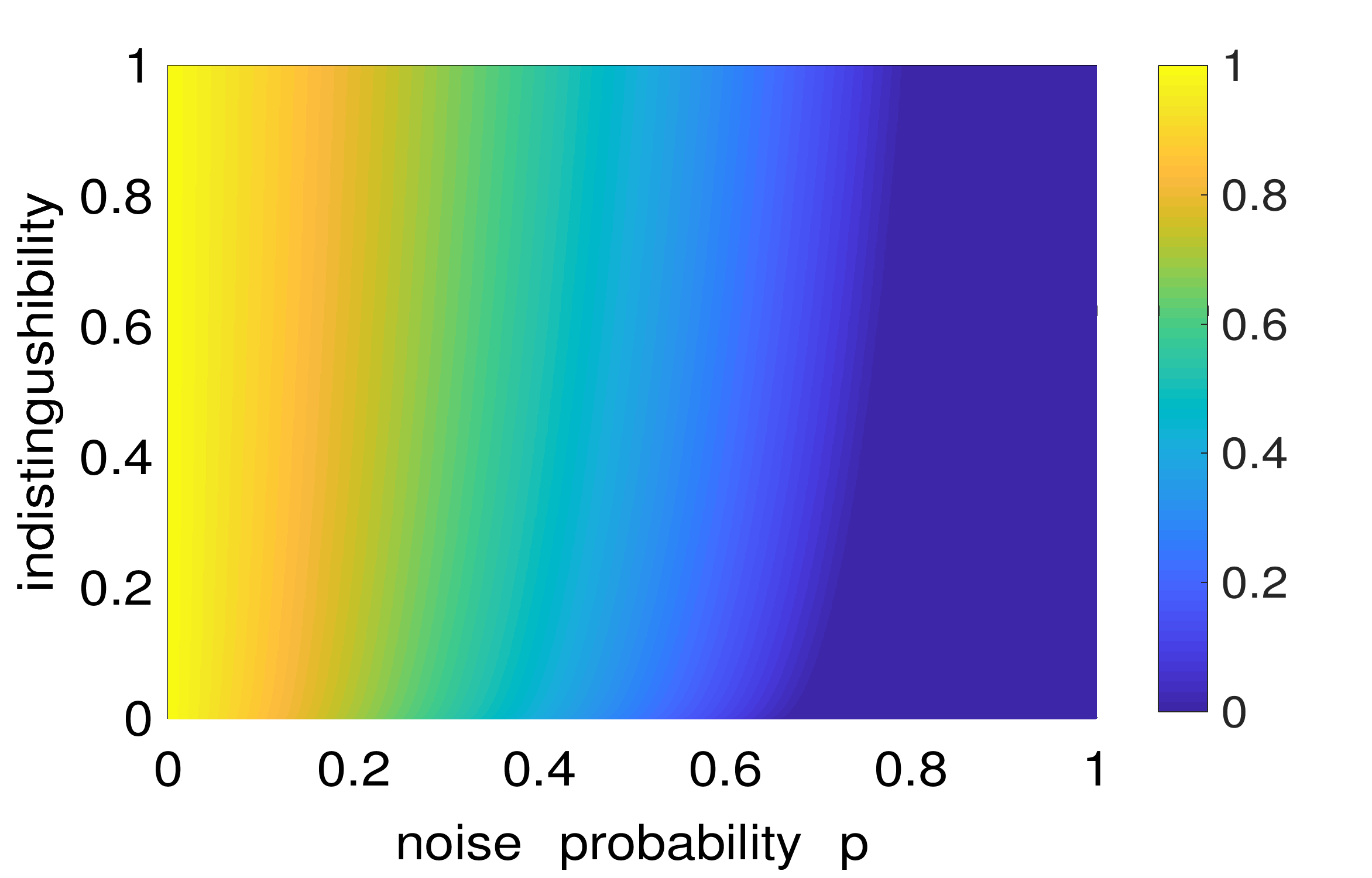}
\caption{Contour plot of entanglement versus noise probability $p$ and spatial indistinguishability $\mathcal{I}_\mathrm{LR}$ when the pure state $\ket{1_{+}}$ is chosen in Eq. \eqref{WindSI}, for fermions (with $\theta=\pi$) or bosons (with $\theta=0$), fixing $l=r'$ in the spatial wave functions of Eq. \eqref{spatialWF}.}
\label{fig:conc2}
\end{figure}

Differently, when the pure state $\ket{1_+}$ is chosen in the $\mathcal{W}^{+}$ state of Eq. \eqref{WindSI}, for fermions (bosons) with $\theta=\pi$ ($\theta=0$) we obtain
\begin{gather}\label{conc2}
   C(\mathcal{W}^{+}_\mathrm{LR})=\max\left\{0,\frac{(4-5 p)(lr'+l'r)^2-p(lr'-l'r)^2}{4\left(l^2 r'^2+l'^2r^2+lr'rl'(2-p)\right)}\right\},
\end{gather}
with probability
\begin{equation}\label{ProbLR1plus}
    P_\mathrm{LR}=\frac{2\left(l^2 r'^2+l'^2r^2+lr'rl'(2-p)\right)}{2
    +\eta (2-p)(ll' +\eta rr')^2}.
\end{equation}
The concurrence of Eq. \eqref{conc2} is plotted in Fig. \ref{fig:conc2} in terms of both indistinguishability degree $\mathcal{I}_\mathrm{LR}$ and noise probability $p$, fixing $l=r'$. By choosing maximal indistinguishability $\mathcal{I}_\mathrm{LR}=1$ ($l=l'$), one has $C(\mathcal{W}^{+}_\mathrm{LR})= (4-5p)/(4-p)$ when $0 \leq p < 4/5$, being zero elsewhere. The sLOCC probability is maximized when $l=l'=1/\sqrt{2}$, taking the expression $P_\mathrm{LR}=1-p/4$ for fermions and $P_\mathrm{LR}=1/2$ for bosons. 
The probability $P_\mathrm{LR}$ of Eq. \eqref{ProbLR1plus} is plotted in Fig. \ref{fig:Pr}(b) as a function of noise probability $p$ for some degrees of spatial indistinguishability $\mathcal{I}_\mathrm{LR}$. 

Finally, the amount of entanglement for a Werner state of distinguishable qubits $W^{\pm}_\mathrm{AB}$ can be retrieved by both Eqs. \eqref{conc1} and \eqref{conc2} of $C(\mathcal{W}^{\pm}_\mathrm{LR})$ for spatially separated wave functions ($\mathcal{I}_\mathrm{LR}=0$). This is reached when $l=r'=1$ or $l=r'=0$ in Eq. \eqref{spatialWF}. In this case, in fact, one can associate $\mathrm{A}\equiv\mathrm{L}$ and $\mathrm{B}\equiv\mathrm{R}$. We thus obtain the known result \cite{horo2009} about the concurrence, that is $C(W^{\pm}_\mathrm{AB})=1-3p/2$ for $ 0 <p \leq 2/3$, being zero elsewhere.

\begin{figure}[b!] 
{\includegraphics[scale=0.23]{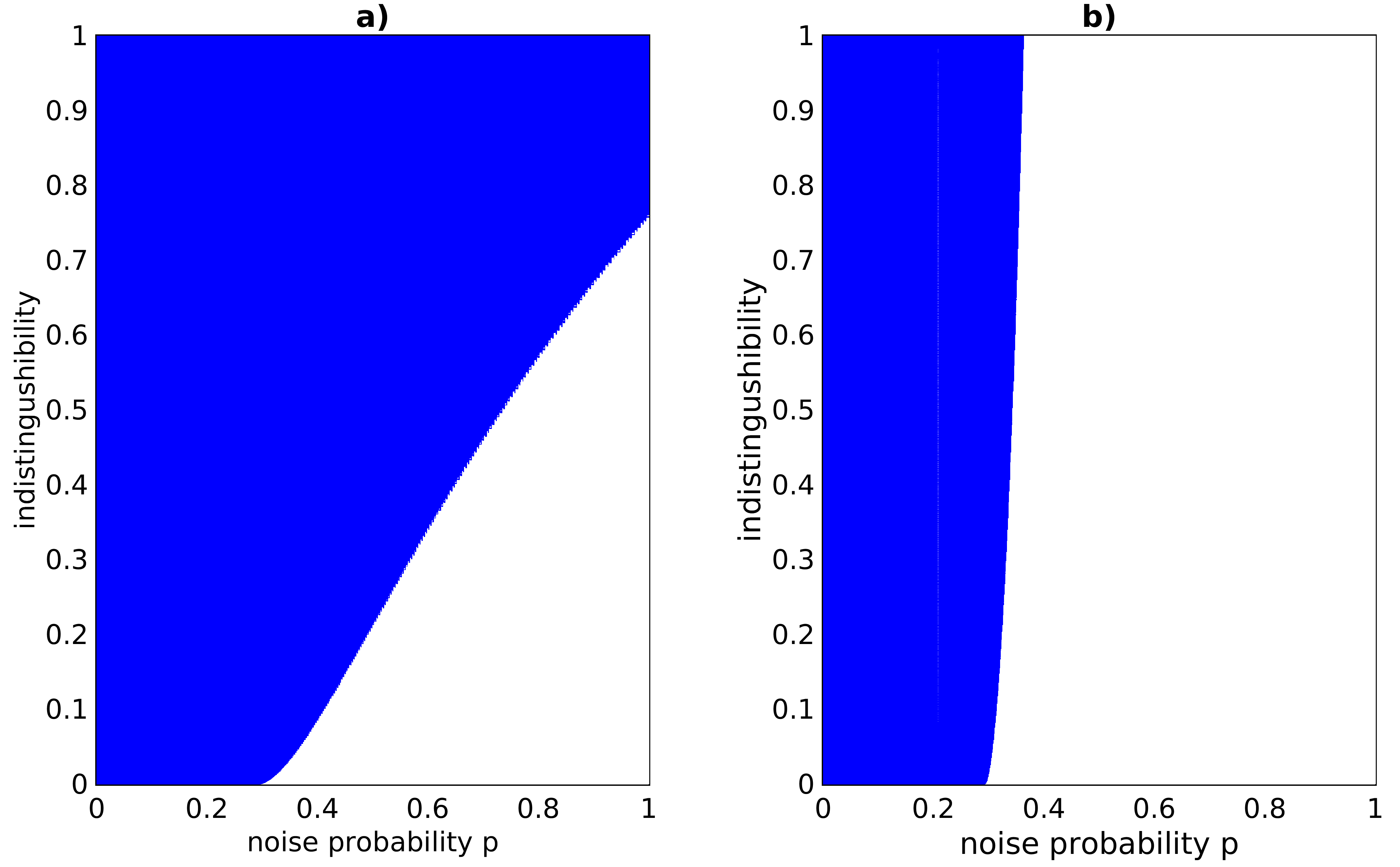}}
\caption{Regions of Bell inequality violation $B(\mathcal{W}^{\pm}_\mathrm{LR})>2$ in terms of noise probability $p$ and degree of spatial indistinguishability $\mathcal{I}_\mathrm{LR}$. To vary $\mathcal{I}_\mathrm{LR}$ we fix $l=r'$ in Eq. \eqref{spatialWF}. \textbf{a.} Target pure state $\ket{1_{-}}$ in Eq. \eqref{WindSI} for both fermions (with $\theta=0$) and bosons (with $\theta=\pi$). \textbf{b.} Target pure state $\ket{1_{+}}$ in Eq. \eqref{WindSI} for both fermions (with $\theta=\pi$) and bosons (with $\theta=0$).}
\label{fig:bell1}
\end{figure}

\section{CHSH-Bell inequality violation for the $\mathcal{W}^\pm$ state}\label{AppD}

\begin{figure*}[t] 
\includegraphics[width=0.75\textwidth]{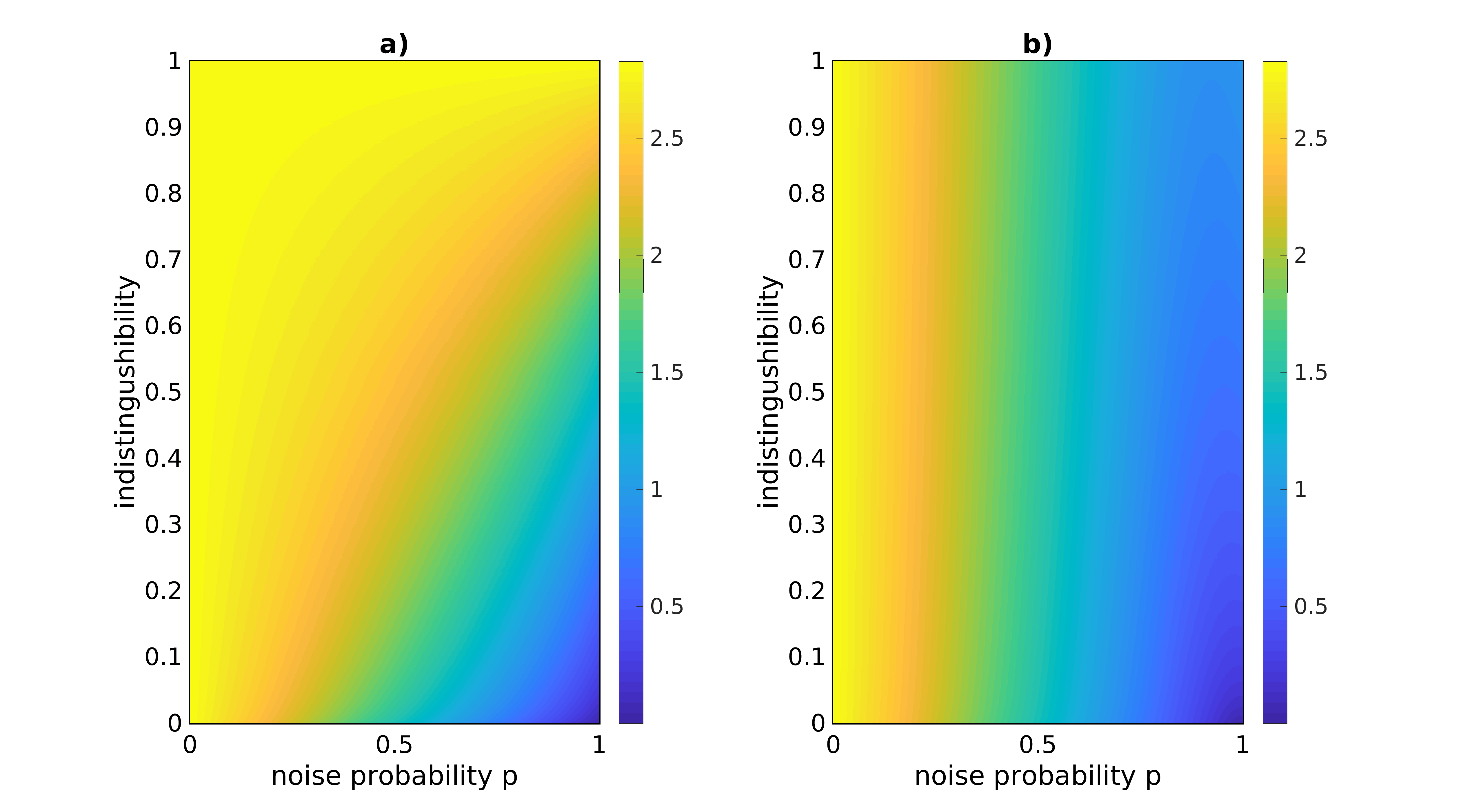}
\caption{Contour plot of Bell function $B(\mathcal{W}^{\pm}_\mathrm{LR})$ versus noise probability $p$ and degree of spatial indistinguishability. To vary $\mathcal{I}_\mathrm{LR}$ we fix $l=r'$ in Eq. \eqref{spatialWF}. \textbf{a.} Target pure state $\ket{1_{-}}$ in Eq. \eqref{WindSI} for both fermions (with $\theta=0$) and bosons (with $\theta=\pi$). \textbf{b.} Target pure state $\ket{1_{+}}$ in Eq. \eqref{WindSI} for both fermions (with $\theta=\pi$) and bosons (with 
$\theta=0$).}
\label{fig:bell2}
\end{figure*}

It is known that for mixed states of distinguishable particles, which are the ones encountered in practice, a given value of entanglement by itself does not guarantee that the correlations cannot be reproduced by a classical local hidden variable model \cite{werner1989quantum,horo2009}. Criteria based on Bell inequality violations are therefore those utilized to show that a given amount of entanglement of mixed states assures nonlocal quantum correlations, which are not classically reproducible \cite{horo2009}. For two distinguishable qubits in an arbitrary state $\rho$, the experimentally-friendly CHSH-Bell inequality can be written as $B(\rho)\leq 2$ \cite{horo2009}, where $2$ represents the classical threshold. Whenever a quantum state produces $B(\rho)> 2$, its corresponding entanglement is inherently nonlocal. 

Thanks to the sLOCC framework, we can straightforwardly translate these arguments to the case of arbitrary mixed states of indistinguishable particles. In fact, after sLOCC the identical qubits can be individually addressed and a Bell test can be performed on their global state by spin-like measurements onto the separated bound states 
$\ket{\mathrm{L}}$, localized in the region $\mathcal{L}$, and $\ket{\mathrm{R}}$, localized in the region $\mathcal{R}$ \cite{PhysRevLett.120.240403}. 

It is simple to see that the distributed resource state $\mathcal{W}^{\pm}_\mathrm{LR}$, stemming from $\mathcal{W}^\pm$ of Eq. \eqref{WindSI} after sLOCC, has an X structure in the computational basis $\mathcal{B}_\mathrm{LR}=\{\ket{\mathrm{L}\uparrow,\mathrm{R}\uparrow}, \ket{\mathrm{L}\uparrow,\mathrm{R}\downarrow}, \ket{\mathrm{L}\downarrow,\mathrm{R}\uparrow}, \ket{\mathrm{L}\downarrow,\mathrm{R}\downarrow}\}$. That is, only the diagonal and off-diagonal elements are in general nonzero. According to the Horodecki criterion for the CHSH-Bell inequality violation of a two-qubit density matrix  \cite{horodecki1995violating}, the expression of the optimized Bell function for an X-shape density matrix $\rho_X$ can be written as \cite{PhysRevA.78.062309}
\begin{equation}
B(\rho_X)=2\sqrt{\mathcal{P}^2+\mathcal{Q}^2},
\end{equation}
with
\begin{gather}
\mathcal{P}=\rho_{11}+\rho_{44}-\rho_{22}-\rho_{33},\quad
\mathcal{Q}=2(|\rho_{14}|+|\rho_{23}|),
\end{gather}
where $\rho_{ij}$ are the density matrix elements in the computational basis. For the state of our interest, $\mathcal{W}^{\pm}_\mathrm{LR}$, these elements are functions of both the degree of spatial indistinguishability $\mathcal{I}_\mathrm{LR}$, through the wave function parameters $l$ and $l'$, and the noise probability $p$. It is thus possible to look for Bell inequality violations, that is $B_\mathrm{LR}(\mathcal{W}^{\pm}):=B(\mathcal{W}^{\pm}_\mathrm{LR})>2$, for different values of $\mathcal{I}_\mathrm{LR}$ and $p$. 
To continuously vary the degree of indistinguishability and forbid that the sLOCC probability $P_\mathrm{LR}$ to get $\mathcal{W}^{\pm}_\mathrm{LR}$ is zero, we fix $l=r'$ in the wave functions $\ket{\psi_1}$ and $\ket{\psi_2}$ of Eq. \eqref{spatialWF}. It is worth to recall that a Bell inequality violation based on sLOCC, that is based on local postselection with $P_\mathrm{LR}>0$, provides a faithful test of local realism \cite{sciarrinoPRA}.

When the target pure state in Eq. \eqref{WindSI} is $\ket{1_-}$, for both fermions (with $\theta=0$) and bosons (with $\theta=\pi$), the behavior of the Bell function $B(\mathcal{W}^{-}_\mathrm{LR})$ is displayed in Fig. \ref{fig:bell1}\textbf{a}, where only the regions of violation are evidenced, while in Fig. \ref{fig:bell2}\textbf{a} all the possible values of $B(\mathcal{W}^{-}_\mathrm{LR})$ are shown. From Fig. \ref{fig:bell1}\textbf{a}, it is clear that when $0.76< \mathcal{I}_\mathrm{LR} \leq 1 $, that is $0.56< C(\mathcal{W}^{-}_\mathrm{LR}) \leq 1 $, the Bell inequality is violated independently of noise probability $p$. The fact that this result is independent of $p$ implies that, under the conditions above, the nonlocal entanglement preparation is always noise free. Moreover, notice that the degree of $\mathcal{I}_\mathrm{LR}$ beyond the (nonlocality) threshold ($\approx 0.76$) is reached for many different shapes of the spatial wave functions of Eq. \eqref{spatialWF}. Disturbance in the control of the exact shape of spatial wave functions does not significantly affect the degree of spatial indistinguishability of the identical qubits. 

On the other hand, when the target pure state in Eq. \eqref{WindSI} is $\ket{1_+}$, for both fermions (with $\theta=\pi$) and bosons (with $\theta=0$), the behavior of $B_\mathrm{max}(\mathcal{W}^{+}_\mathrm{LR})$ is reported in Fig. \ref{fig:bell1}\textbf{b}, where only the regions of violation are evidenced, while in Fig. \ref{fig:bell2}\textbf{b} all the possible values of the Bell function are plotted. In this case, Fig. \ref{fig:bell1}\textbf{b} evidences that the Bell inequality, also for a high degree of indistinguishability, can be violated for a finite range of $p$. For instance, when $\mathcal{I}_\mathrm{LR}=1$, one has $B(\mathcal{W}^{+}_\mathrm{LR})>2$ for $0 \leq p < 0.363$, which means $ 0.6 <  C(\mathcal{W}^{+}_\mathrm{LR}) \leq 1$. 

Finally, from both panels of Fig. \ref{fig:bell1} one can also retrieve the range of values of $p$ for which the Bell inequality is violated in the case of distinguishable qubits A, B. In fact, when $\mathcal{I}_\mathrm{LR}=0$, one gets 
$B(W^{\pm}_\mathrm{AB})>2$ for noise probability 
$0 \leq p<0.29$, which means an entanglement amount $0.68 < C(W^{\pm}_\mathrm{AB}) \leq 1$ \cite{horo2009}. Nonlocal entanglement of distinguishable particles is unavoidably affected by the amount of noise present in the state and is violated only for a small amount of white noise. 

The main message of this analysis is the following: spatial indistinguishability, even partial, can guarantee a noise-free preparation of highly entangled states of identical particles, capable to violate the Bell inequality. Moreover, this result is robust to changes in the configurations of the spatial wave functions $\ket{\psi_1}$, $\ket{\psi_2}$, which highlights the behavior of quantum indistinguishability as a protection of quantum states against external noise.


\end{document}